\documentclass[a4paper,11pt]{article}
\usepackage{pos}

\usepackage{amsfonts} 
\usepackage{amsmath} 
\usepackage{subcaption}
\usepackage{array} 
\usepackage{bm} 
\usepackage{latexsym} 
\usepackage{longtable} 
\usepackage{hyperref} 
\usepackage{multirow}
\usepackage{xcolor}
\usepackage{bbold}
\usepackage{lipsum}
\usepackage{mathtools}
\usepackage{comment}
\usepackage{paralist}
\usepackage{placeins}
\usepackage{verbatim}
\usepackage{adjustbox}

\graphicspath{{./figs/}}

\allowdisplaybreaks

\definecolor{HLBlue}{HTML}{6599FF}
\definecolor{HLOrange}{HTML}{FF6600}

\newcommand{\Vcb}{|V_{cb}|}

\newcommand{\BtoDst}{\bar{B} \to D^\ast \ell \bar{\nu}}
\newcommand{\BtoD}{\bar{B} \to D \ell \bar{\nu}}

\title{Improved data analysis on two-point correlation function with
  sequential Bayesian method}
\ShortTitle{The Newton method}

\author[a,1]{Tanmoy Bhattacharya}
\author*[b,1]{Benjamin J.~Choi}
\author[a,1]{Rajan Gupta}
\author[c,1]{Yong-Chull Jang}
\author[b,1]{Seungyeob Jwa}
\author[b,1]{Sunkyu Lee}
\author[b,1]{Weonjong Lee}
\author[d,1]{Jaehoon Leem}
\author[e,1]{Sungwoo Park}
\author[f,1]{Boram Yoon}

\affiliation[a]{Theoretical Division T-2, Los Alamos National
  Laboratory, Los Alamos, NM 87545, USA}

\affiliation[b]{Lattice Gauge Theory Research Center, CTP and FPRD,
  Department of Physics and Astronomy,\\ Seoul National University,
  Seoul 08826, South Korea}

\affiliation[c]{Department of Physics, Columbia University, 538 West
  120th Street, New York, NY 10027, USA}

\affiliation[d]{School of Physics, Korea Institute for Advanced Study
  (KIAS), Seoul 02455, South Korea}

\affiliation[e]{Thomas Jefferson National Accelerator Facility, 12000
  Jefferson Avenue, Newport News, VA 23606, USA}

\affiliation[f]{Computer, Computational and Statistical Science
  Division CCS-7, Los Alamos National Laboratory, Los Alamos, NM
  87545, USA}

\note{For the LANL-SWME Collaboration}

\emailAdd{benjaminchoi@snu.ac.kr}

\emailAdd{wlee@snu.ac.kr}

\abstract{ We report our progress in data analysis on two-point
  correlation functions of the $B$ meson using sequential Bayesian
  method.
  The data set of measurement is obtained using the Oktay-Kronfeld (OK)
  action for the bottom quarks (valence quarks) and the HISQ action
  for the light quarks on the MILC HISQ lattices.
  We find that the old initial guess for the $\chi^2$ minimizer in the
  fitting code is poor enough to slow down the analysis somewhat.
  In order to find a better initial guess, we adopt the Newton method.
  We find that the Newton method provides a natural test to check
  whether the $\chi^2$ minimizer finds a local minimum or the global
  minimum, and it also reduces the number of iterations
  dramatically. }

\FullConference{ The 38th International Symposium on Lattice Field
  Theory, LATTICE2021 26th-30th July, 2021 Zoom/Gather@Massachusetts
  Institute of Technology }

\begin{document}
\maketitle

\section{Introduction}
\label{sec:intr}

To determine Cabibbo-Kobayashi-Maskawa (CKM) matrix element $\Vcb$, we
need to calculate the semileptonic form factors for the $\BtoDst$ and
$\BtoD$ decays on the lattice.
In order to obtain the semileptonic form factors, we need to do the
data analysis on the three-point (3pt) correlation functions, which
require results for the masses and normalization constants obtained
from the two-point (2pt) correlation functions.
Here, we present recent progress in our data analysis on the 2pt
correlation functions to determine the masses and normalization
constants for the ground and excited states.

We adopt the Fermilab formulation \cite{ElKhadra:1996mp} to implement
the heavy quarks such as bottom and charm quarks on the lattice.
The Fermilab action \cite{ElKhadra:1996mp} is improved up to the
$\lambda^1$ level ($\lambda \simeq \Lambda/(2 m_Q)$), and so it is
impossible to achieve a sub-percent precision with it by construction.
In order to overcome this difficulty, we use the Oktay-Kronfeld (OK)
action \cite{Oktay:2008ex} improved up to the $\lambda^3$ level.
Recently, we have completed the current improvement up to the same
level as the OK action \cite{Bailey:2020uon}.
The OK action for a heavy quark is
\begin{align}
  S_{\text{OK}} & = S_0 + S_1 + S_2 + S_3
  \label{eq:ok-ac-1}
\end{align}
where $S_n$ represents those $\mathcal{O}(\lambda^n)$ terms of
$S_{\text{OK}}$ collectively.
\begin{align}
  S_0 & = a^{4} \sum_{x} \bar{\psi} (x) \left[ m_{0} + \gamma_{4}
    D_{4} \right] \psi(x) \,, \nonumber \\
  S_1 & = a^{4} \sum_{x} \bar{\psi} (x) \left[ - \frac{1}{2} a
    \Delta_{4} + \zeta \bm{\gamma \cdot D} - \frac{1}{2} r_{s} \zeta a
    \Delta^{(3)} - \frac{1}{2} c_{B} a \zeta i \bm{\Sigma \cdot B}
    \right] \psi(x) \,, \nonumber \\
  S_2 & = a^{4} \sum_{x} \bar{\psi} (x) \left[ - \frac{1}{2} c_{E} a
    \zeta \bm{\alpha \cdot E}
    \right] \psi(x) \,, \nonumber \\
  S_3 & = a^{4} \sum_{x} \bar{\psi} (x) \left[ c_{1} a^{2} \sum_{k}
    \gamma_{k} D_{k} \Delta_{k} + c_{2} a^{2} \{ \bm{\gamma \cdot
      D},\Delta^{(3)} \} + c_{3} a^{2} \{ \bm{\gamma \cdot D} \, , i
    \bm{\Sigma \cdot B} \} \right. \nonumber \\
    & \hphantom{ = a^{4} \sum_{x} \bar{\psi} (x) [\ } \left. + c_{EE}
      a^{2} \{ \gamma_{4} D_{4} \, , \bm{\alpha \cdot E} \} + c_{4}
      a^{3} \sum_{k} \Delta_{k}^{2} + c_{5} a^{3} \sum_{k} \sum_{j
        \neq k} \{ i \Sigma_{k} B_{k} \, , \Delta_{j} \} \right]
    \psi(x) \,.
    \label{eq:ok-ac-2}    
\end{align}
Here, we use the same notation as in Ref.~\cite{Oktay:2008ex}.
The bare quark mass $m_0$ is 
\begin{align}
  a m_{0} = \frac{1}{2} \left( \frac{1}{\kappa} -
  \frac{1}{\kappa_{\text{crit}}} \right)\,,
  \label{eq:bare-m-1}
\end{align}
where $\kappa$ ($\kappa_\text{crit}$) is a (critical) hopping
parameter \cite{Bailey:2017xjk}.
Numerical values for $\kappa$ and $\kappa_\text{crit}$ are
summarized in Table \ref{tab:parameters}.
We use the HISQ action \cite{Follana:2006rc} for strange quarks.

In order to get a better signal for the $B_{s}$ meson states, we apply
a covariant Gaussian smearing (CGS), $\left\{1 + \sigma^2\nabla^2/(4
N_{\text{GS}}) \right\}^{N_{\text{GS}}}$ to the point source and sink
as in Ref.~\cite{Yoon:2016dij}.
The CGS parameters are summarized in Table \ref{tab:parameters}.
We apply the CGS only to the heavy quark fields and not to the light
quark fields.
\begin{table}[t!]
  \renewcommand{\arraystretch}{1.2}
  \center
  \resizebox{1.0\textwidth}{!}{
    \begin{tabular}{@{\qquad} c @{\qquad}|@{\qquad} c @{\qquad}@{\qquad} c @{\qquad}|@{\qquad} c @{\qquad}|@{\qquad} c @{\qquad}}
      \hline\hline
      $m_x / m_s$ & $\kappa_{\text{crit}}$ & $\kappa_b$ & $\{ \sigma
      \,,\; N_{\text{GS}} \}$ & $N_{\text{cfg}} \times N_{\text{src}}$
      \\
      \hline
      1 & 0.051218 & 0.04070 & $\{ 1.5\,,\;5 \}$ & $1000 \times 3 $ \\
      \hline\hline
    \end{tabular}
  } 
  \caption{ Hopping parameters and smearing parameters.  Here, $m_x$
    is a mass of valence light quarks.  $N_{\text{cfg}}$ represents
    the number of gauge configurations and $N_{\text{src}}$ is the
    number of sources per gauge configuration.  }
  \label{tab:parameters}
\end{table}

We use the MILC HISQ ensembles with $N_f=2+1+1$ \cite{
  Bazavov:2012xda}.
The details are summarized in Table \ref{tab:ensembles}.
\begin{table}[!t]
  \renewcommand{\arraystretch}{1.2}
  \center
  \resizebox{1.0\textwidth}{!}{
    \begin{tabular}{ @{\qquad} c @{\qquad}|@{\qquad} c @{\qquad}|@{\qquad} c @{\qquad}|@{\qquad} c @{\qquad} c @{\qquad} c @{\qquad}}
    \hline\hline $a$ (fm) & $N_s^3 \times N_t$ & $M_\pi$ (MeV) &
    $am_l$ & $am_s$ & $am_c$ \\
    \hline
    0.1184(10) & $32^3\times 64$ & 216.9(2) & 0.00507 & 0.0507 & 0.628
    \\
    \hline\hline
    \end{tabular}
  } 
  \caption{ Details on the MILC HISQ ensembles with $N_f=2+1+1$
    \cite{Bazavov:2012xda}.}
  \label{tab:ensembles}
\end{table}

%
%

%
\section{Sequential Bayesian Method}
\label{sec:corfit}

Let us consider 2pt correlation functions \cite{ Bazavov:2011aa}:
\begin{align}
  C(t) & = \sum_{\alpha = 1}^{4} \sum_{\bf{x}} \langle
  \mathcal{O}^{\dagger}_\alpha(t,{\bf x}) \mathcal{O}_\alpha(0)
  \rangle
  \label{eq:corr-2pt}
\end{align}
Here, the interpolating operator for the heavy-light meson is
\begin{align}
  \mathcal{O}_\alpha(t,{\bf x}) &= \left[ \bar{\psi}_b(t,{\bf x})
    \gamma_5 \Omega(t,{\bf x}) \right]_\alpha \chi_\ell(t,{\bf x}) \,.
\end{align}
Here $\psi_b$ is an OK action field for bottom quarks, and
$\chi_\ell$ is an HISQ action field for light quarks.
\begin{align}
\Omega(t,{\bf x}) \equiv \gamma_{1}^{\;x_1} \gamma_{2}^{\;x_2}
\gamma_{3}^{\;x_3} \gamma_{4}^{\;t} \,.
\label{eq:omega-trans-1}
\end{align}
The subscript $\alpha$ represents taste degrees of freedom for
staggered light quarks.
We construct the fitting function to contain $m$ even time-parity
states and $n$ odd time-parity states, which we call
``$m+n$ fit''.
The $m+n$ fit function is
\begin{align}
  f(t) & = g(t) + g(T-t)\,, \nonumber \\
  g(t) &= A_0 \, e^{-E_0 \, t} \left[ 1 + R_{2} \, e^{-\Delta E_{2} \,
      t} \left( 1 + R_{4} \, e^{-\Delta E_{4} \, t} \left( \cdots
    \left( 1 + R_{2m-2} \, e^{-\Delta E_{2m-2} \, t} \right) \cdots
    \right) \right) \right. \nonumber \\
    & \hphantom{= A_0 \, e^{-E_0 \, t}[ \ } \left. - (-1)^{t} \, R_{1}
      \, e^{-\Delta E_{1} \, t} \left( 1 + R_{3} \, e^{-\Delta E_{3}
        \, t} \left( \cdots \left( 1 + R_{2n-1} \, e^{-\Delta E_{2n-1}
        \, t} \right) \cdots \right) \right) \right]
  \label{eq:fit-func}
\end{align}
where $\Delta E_i \equiv E_i - E_{i-2}$, $E_{-1}\equiv E_0$\,, $R_i
\equiv { A_i }/{ A_{i-2} }$ and $A_{-1} \equiv A_0$.

We adopt the sequential Bayesian method for fitting.
We take the following steps to analyze the 2-point correlation
functions.
\begin{description}
\item [Step 1] Do the 1st fitting. \quad
  ex) 1+0 fit (2 parameters: \{$A_0$, $E_0$\})
\item [Step 2] Feed the fitting results as prior information for the 2nd
  fitting. \\
  ex) 1+1 fit (4 parameters: \{$A_0$, $E_0$, $R_1$, $\Delta
  E_1$\}, 2 prior information on \{$A_0$, $E_0$\})
\item [Step 3] Do stability test and find optimal prior
  information.
  ex) stability test gives optimal prior information on
  \{$A_0$, $E_0$\}.
\item [Step 4] Save the 2nd fitting results (\textit{e.g.}~1+1 fit)
  into the 1st fitting.
\item [Step 5] Choose the next fitting (\textit{e.g.}~2+1 fit) as the
  2nd fitting.
\item [Step 6] Go back to \textbf{Step 2}. \quad
  ex) 1+0 $\to$ 1+1 $\to$ 2+1 $\to$ 2+2 $\to$ $\cdots$.
\end{description}

%
%

%
\section{Numerical precision problem on covariance matrix inversion}
\label{sec:covmat-inv}

When we fit the data to the fitting function given in
Eq.~\eqref{eq:fit-func}, we encounter a numerical precision
problem in the covariance matrix inversion.
For example, we set the fitting range to $15 \le t \le 29$ and then
we have a covariance matrix $V$ of $15 \times 15$.
We use the Cholesky decomposition algorithm to obtain the inverse
matrix $V^{-1}$.
In order to check the matrix inversion, we monitor the following
identity: $V \cdot V^{-1} = 1$.
If everything works well, we will get the off-diagonal components of
$V \cdot V^{-1}$ to be zero within numerical precision, but we find
that some of them are $\mathcal{O}(10^{-5})$.
We also find that the largest and smallest eigenvalues for $V$ are
$\mathcal{O}(10^{-35})$ and $\mathcal{O}(10^{-60})$, respectively.
Since $V^{-1}$ is used multiple times in the least $\chi^2$ fitting,
we need $V^{-1}$ accurate up to double precision, but the existing
fitting code cannot achieve the numerical precision.
Hence, we find two independent methods to resolve the puzzle: one is
the rescaling method and the other the correlation matrix method.

\subsection{Rescaling method}
\label{sec:rsc-1}

We transform the data $C(t)$, the fit function $f(t)$ and the
covariance matrix $V(t_i,t_j)$ by an \textit{arbitrary} rescaling
factor $R(t)$ as follows,
\begin{align}
  \tilde{C}(t) & = \frac{C(t)}{R(t)} \,, \quad
  \tilde{f}(t) = \frac{f(t)}{R(t)} \,, \quad
  \tilde{V}(t_i,t_j) = \frac{V(t_i,t_j)}{R(t_i)\,R(t_j)} \,.
  \label{eq:res-trans}
\end{align}
Then the fitting results and the $\chi^2$ value are invariant under
the rescaling transformation of Eqs.~\eqref{eq:res-trans},
regardless of details on $R(t)$.
Here, note that the fitting parameters never be changed by the rescaling
factor $R(t)$.

In our data analysis, we set the rescaling function to
\begin{align}
  R(t) &= A_0^{r} \exp[-E_0^{r} t] + A_0^{r} \exp[-E_0^{r} ( T - t ) ] \,.
  \label{eq:R}
\end{align}
Here, $A_0^r$ and $E_0^r$ is determined by fitting the data in the fit
range ($23 \le t \le 29$), where the superscript $r$ represents the
rescaling function.
The huge scale difference between the largest and the smallest
eigenvalues of $V$ comes from the large mass ($E_0 \simeq 2.0/a$) of
the $B_s$ meson, since the 2pt correlation function decreases as
a function of $\sim \exp(-E_0 t)$ at the leading order.
Hence, if we remove this leading order exponential decay term by the
rescaling function $R(t)$ in Eq.~\eqref{eq:R}, then the remaining
scale difference in the largest and the smallest eigenvalues of
$\tilde{V}$ reduces to the $\mathcal{O}(10^{-2})$ level, which allows
us to use the Cholesky algorithm reliably for the matrix inversion.
We find that the off-diagonal components of $ \tilde{V} \cdot
\tilde{V}^{-1}$ are zero within the numerical precision.
Therefore, the rescaling method resolves our numerical precision
problem.
%

\subsection{Correlation matrix method}
\label{sec:corrmat-1}

For a given covariance matrix $V(t_i,t_j)$, we define the correlation
matrix as
\begin{align}
  \rho(t_i,t_j) &\equiv \frac{V(t_i,t_j)}{\sigma(t_i) \sigma(t_j)} \quad
  \text{where} \quad \sigma(t_i) = \sqrt{V(t_i,t_i)} \,,
\end{align}
Then we can obtain the inverse covariance matrix $V^{-1}$ using
the following simple identity:
\begin{align}
  V^{-1}(t_i,t_j) & = \frac{\rho^{-1}(t_i,t_j) }{ \sigma(t_i) \,
    \sigma(t_j)} \,.
\end{align}
Here, note that the correlation matrix $\rho(t_i,t_j)$ is
$\mathcal{O}(1)$, while $\sigma(t)$ decays exponentially like the
rescaling function $R(t)$ in the previous subsection.
The remaining scale difference in the largest and smallest eigenvalues
of $\rho$ reduces to the $10^{-2}$ level.
Hence, the correlation matrix method also resolves our numerical
precision problem.

\subsection{Comparison of the rescaling and correlation matrix methods}
\label{sec:comparison-1}

In Table \ref{tab:comparison-1}, we present the fitting results obtained
using the rescaling method and the correlation matrix method.
We find that both provides the same results.
The difference in computing time is negligible (only $\simeq 0.7\%$).
We conclude that both methods are good for our fitting purpose.
Hence, we use both methods here to crosscheck the fitting results
by comparison.
\begin{table}[h!]
  \renewcommand{\arraystretch}{1.2}
  \center
  \resizebox{1.0\textwidth}{!}{
    \begin{tabular}{ @{\qquad} c @{\qquad\qquad}|@{\qquad\qquad} l @{\qquad\qquad}|@{\qquad\qquad} l @{\qquad}}
      \hline
      \hline
      parameter & rescaling & correlation \\
      \hline
      $A_{0}$ & 0.01724(52) & 0.01724(52) \\
      $E_{0}$ & 2.0448(22) & 2.0448(22) \\
      $R_{1}$ & 3.5(58) & 3.5(58) \\
      $\Delta E_{1}$ & 0.36(12) & 0.36(12) \\
      \hline
      $\chi^2/\text{d.o.f.}$ & 0.2306(80) & 0.2306(80) \\
      computing time [sec] & 73.3 & 72.8 \\
      \hline
      \hline
    \end{tabular}
  } 
  \caption{Comparison of the rescaling method and the correlation
    method for the 1+1 fit with the fit range of $13 \le t \le 29$.}
  \label{tab:comparison-1}
\end{table}

%

%
%

%

\section{ Application of the Newton method to the initial guess for
  the $\chi^2$ minimizer }
\label{sec:newton-initial}

When we do the least $\chi^2$ fitting, we use the
Broyden-Fletcher-Goldfarb-Shanno (BFGS) algorithm \cite{
  Broyden:1970bro, Fletcher:1970fle, Goldfarb:1970gol, Shanno:1970sha}
for the $\chi^2$ minimizer.
The BFGS algorithm is one of the quasi-Newton methods for minimization.
The BFGS algorithm needs an initial guess for the fitting parameters
by construction. 
The old version of our fitting code sets up the initial guess as
follows.
First, solve Eq.~\eqref{eq:init-guess-1} to obtain an initial guess
for $A_0$ and $E_0$.
\begin{align}
  \begin{pmatrix}
    \displaystyle\sum_i \, \dfrac{C^2(t_i)}{\sigma^2(t_i)} &
    \displaystyle\sum_i \, t_i \, \dfrac{C^2(t_i)}{\sigma^2(t_i)} \\
    \displaystyle\sum_i \, t_i \, \dfrac{C^2(t_i)}{\sigma^2(t_i)} &
    \displaystyle\sum_i \, t_i^2 \, \dfrac{C^2(t_i)}{\sigma^2(t_i)}
  \end{pmatrix}
  \begin{pmatrix}
    \ln A_{0}^g \\
    -E_{0}^g
  \end{pmatrix}
  & =
  \begin{pmatrix}
    \displaystyle\sum_i \, \dfrac{C^2(t_i)}{\sigma^2(t_i)} \, \ln
    \left| C(t_i) \right| \\
    \displaystyle\sum_i \, t_i \, \dfrac{C^2(t_i)}{\sigma^2(t_i)} \,
    \ln \left| C(t_i) \right|
  \end{pmatrix}
  \label{eq:init-guess-1}
\end{align}
where the superscipt $g$ in $A_0^g$ and $E_0^g$ represents the initial
guess.
Second, in order to obtain an initial guess for $R_i$ and $\Delta
E_i$, the old fitting code adopts the following convention:
\begin{align}
  R_{2j}^g & = 2.5\,j \,, \quad R_{2j-1}^g = 0.025\,j \,,
  \label{eq:init-guess-2}
  \\
  \Delta E_{2j}^g &= \Delta E_{2j-1}^g = 0.1 E_0^g \,,
  \label{eq:init-guess-3}
\end{align}
where $j \ge 1$ and $j \in Z$.

For example, in the 3+2 fit, the old fitting code sets up the initial
guess to $R_4^g = 5.0$, $\Delta E_4^g = 0.1 \; E_0^g$.
However, we find that $R_i \lesssim 1$ typically in our fitting.
Since the initial guess values for $R_i$ is very far away from the
fitting results for $R_i$, the $\chi^2$ minimizer (a quasi-Newton
method) works too hard to get a realistic value for $R_i$, which is
not necessary, if one can feed a better initial guess for $R_i$ to
the $\chi^2$ minimizer.
In the end of the day, we find that a poor determination of the
initial guess causes the number of iterations for the $\chi^2$
minimizer to increase significantly.

In order to obtain a better initial guess, we use the
multi-dimensional Newton method \cite{ Press:2007nr, Broyden:1965br}.
The Newton method determines the initial guess directly from the
data.
Technical details on the Newton method are described in Subsections
\ref{sec:newton-app}, \ref{sec:deter-A0-E0}, and \ref{sec:scanning}.
In the Table \ref{tab:com-time-1}, we present the number of iterations
for the $\chi^2$ minimizer when we use the old initial guess and the
new initial guess with the Newton method.
Here, we find that the overhead from the Newton method is negligibly
small (about 0.5\% of the running time for a single sample).
\begin{table}[h!]
  \renewcommand{\arraystretch}{1.2}
  \center
  \resizebox{1.0\textwidth}{!}{
    \begin{tabular}{ @{\qquad\qquad} c @{\qquad\qquad}|@{\qquad\qquad} c @{\qquad\qquad}|@{\qquad\qquad} c @{\qquad\qquad}}
      \hline
      \hline
      fit type & old initial guess & Newton method \\
      \hline
      $1+1$ & 1641 & 824 \\
      \hline
      $2+1$ & 1627 & 327 \\
      \hline
      $2+2$ & 1673 & 704 \\
      \hline
      \hline
    \end{tabular}
  } 
  \caption{ Number of iterations of the $\chi^2$ minimizer for a single
    sample. }
  \label{tab:com-time-1}
\end{table}

%

\subsection{The Newton method}
\label{sec:newton-app}
%
When we do the $m+n$ fit, then we have to determine $N=2(m+n)$ fit
parameters.
Hence, we need to choose $N$ time slices such as $\{ t_1, t_2,\ldots,
t_N \}$ in order to apply the multi-dimensional Newton method to find
roots for Eqs.~\eqref{eq:newton-sim-eq-N}.
\begin{align}
  \mathcal{X}(t_i) &\equiv \frac{f(t_i) - C(t_i)}{C(t_i)}
  \label{eq:newton-eq-X-1}
  \\
  \mathcal{X}(t_i) &= 0
  \label{eq:newton-sim-eq-N}
\end{align}
To measure the convergence of the Newton method, we introduce
$\mathbf{D}_N$, the norm of relative difference:
\begin{align}
  {\bf D}_N &= \sqrt{ \sum_{i=1}^{N} \left[
     \mathcal{X}(t_i) \right]^2 } \,,
  \label{eq:rel-diff-1}
\end{align}

To resolve the precision problem in Jacobian matrix inversion, we use
$C(t_i)$'s as rescaling factor in Eq.~\eqref{eq:newton-eq-X-1}.
By rescaling, the Newton method converges faster, while the Jacobian
matrix inversion gets stabilized.
The stopping condition for the Newton method is
\begin{align}
  \max_{i=1,\ldots,N} \, \left| \mathcal{X}(t_i) \right| & < 10^{-12} \,.
\end{align}
%

\subsection{Initial guess for the Newton method in the 1+0 fit}
\label{sec:deter-A0-E0}

We also need an initial guess for the Newton method.
First, we choose two time slices $t_1$ and $t_2 = t_1+2$.
Second, we set the initial guess as follows,
\begin{align}
  E_0^{gn} & = \frac{1}{2} \ln \frac{C(t_1)}{C(t_2)}
  \label{eq:E0-gn}
  \\
  A_0^{gn} & = \frac{C(t_1)}{\exp [ - E_0^{gn} \, t_1 ] + \exp [ -
      E_0^{gn} \, ( T - t_1 ) ]} \,.
  \label{eq:A0-gn}  
\end{align}
Here, the superscript ${}^{gn}$ indicates the initial guess
for the Newton method.
For example, when we set $t_1=21$, we find that ${\bf D}_2 = 7.54
\times 10^{-3}$ for the initial guess, which is good enough
to apply the Newton method to find the exact roots.
%

\subsection{Initial guess for the Newton method for the $1+1$ fit
  with the scanning method}
\label{sec:scanning}

\begin{figure}[h!]
  \centering
  \includegraphics[width=0.6\textwidth]{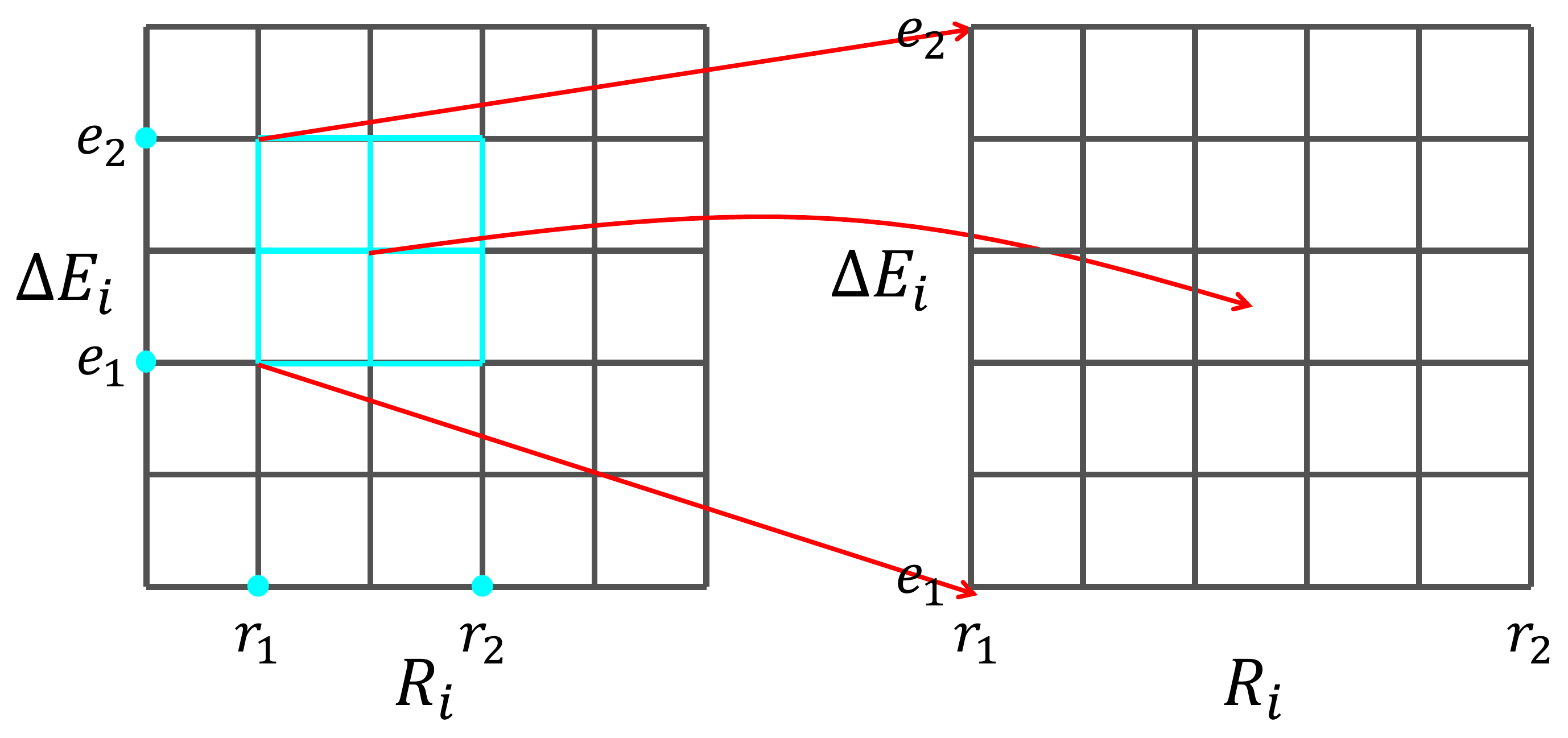}
  \caption{ Schematic picture of iterative scanning}
  \label{fig:iter-scan}
\end{figure}

When we move from one fit to the next fit (\textit{e.g.} 1+0 fit $\to$
1+1 fit), we introduce two or four new fit parameters (\textit{e.g.}
$R_1$ and $\Delta E_1$ for the 1+1 fit) on top of the previous fit
parameters (\textit{e.g.} $A_0$ and $E_0$ for the 1+0 fit), while we
extend the fitting range toward the source time slice.
Here, let us choose the [1+0 $\to$ 1+1] fit as an example to explain
how to set the initial guess for the Newton method.
Since we know the fit results for the 1+0 fit, we may recycle them to
set up the initial guess for $A_0$ and $E_0$.
In order to find an initial guess for the new parameters $R_1$ and
$\Delta E_1$, we use the scanning method as shown in
Fig.~\ref{fig:iter-scan}.
First, we find a proper range for $R_1$ and $\Delta E_1$ such as
$ R_1 \in [0.0, 3.0]$ and $ \Delta E_1 \in [0.0, 1.0]$ and choose
two time slices within the fit range.
Second, we introduce a $6 \times 6$ lattice to cover the full range
as in Fig.~\ref{fig:iter-scan}.
Third, we find the minimum of $\mathbf{D}_2$ on the lattice.
Fourth, we find the new range which contains the nearest neighbor
lattice points of the minimum as in Fig.~\ref{fig:iter-scan}.
Fifth, we repeat the above scanning method until we find $R_1^{gn}$ and
$\Delta E_1^{gn}$ which satisfy the stopping condition $\mathbf{D}_2 <
1.0 \times 10^{-2}$.

%

%
%

%

\section{Results}
\label{sec:corfit-results}

%

\begin{table}[t!]
  \renewcommand{\arraystretch}{1.2}
  \center
  \resizebox{1.0\textwidth}{!}{
    \begin{tabular}{ @{\;} c @{\;}|@{\;} c @{\;\;} c @{\;}|@{\;} c @{\;\;} c @{\;}|@{\;} c @{\;\;} c @{\;}|@{\;} c @{\;\;} c @{\;}|@{\;} c @{\;\;} c @{\;}}
      \hline
      \hline
      fit type & \multicolumn{2}{@{\;}c @{\;}|@{\;}}{$1+0$} & \multicolumn{2}{@{\;}c@{\;}|@{\;}}{$1+1$} & \multicolumn{2}{@{\;}c@{\;}|@{\;}}{$2+1$} & \multicolumn{2}{@{\;}c@{\;}|@{\;}}{$2+2$ (1st)} & \multicolumn{2}{@{\;}c@{\;}}{$2+2$ (2nd)} \\
      \hline
      fit info & Prior & Result & Prior & Result & Prior & Result & Prior & Result & Prior & Result \\ 
      \hline
      $A_0$        & none & 0.0182(29) & 0.018(14) & 0.01724(52) & 0.017(10) & 0.01660(86) & 0.017(10) & 0.01724(35) & 0.017(10) & 0.01727(35) \\
      $E_0$        & none & 2.0468(76) & 2.05(11)  & 2.0448(22)  & 2.045(23) & 2.0428(31)  & 2.043(23) & 2.0449(18)  & 2.045(23) & 2.0450(18) \\
      $R_1$        &      &            & none      & 3.5(58)     & 3.5(35)   & 0.755(82)   & 0.76(76)  & 0.646(84)   & 0.65(34)  & 0.639(79) \\
      $\Delta E_1$ &      &            & none      & 0.36(12)    & 0.36(36)  & 0.255(12)   & 0.26(26)  & 0.242(14)   & 0.24(11)  & 0.241(13) \\
      $R_2$        &      &            &           &             & none      & 0.93(37)    & 0.93(93)  & 1.879(76)   & 1.9(19)   & 1.888(75) \\
      $\Delta E_2$ &      &            &           &             & none      & 0.33(10)    & 0.33(33)  & 0.475(21)   & 0.48(48)  & 0.477(21) \\
      $R_3$        &      &            &           &             &           &             & none      & 2.10(49)    & none      & 2.05(43) \\
      $\Delta E_3$ &      &            &           &             &           &             & none      & 0.58(15)    & none      & 0.57(14) \\
      \hline
      fit range & \multicolumn{2}{@{\;}c @{\;}|@{\;}}{$21 \le t \le 29$} & \multicolumn{2}{@{\;}c @{\;}|@{\;}}{$13 \le t \le 29$} & \multicolumn{2}{@{\;}c @{\;}|@{\;}}{$7 \le t \le 29$} & \multicolumn{2}{@{\;}c @{\;}|@{\;}}{$3 \le t \le 29$} & \multicolumn{2}{@{\;}c @{\;}}{$3 \le t \le 29$} \\
      \hline
      \hline
    \end{tabular}
  } 
  \caption{ Preliminary results from the sequential Bayesian fitting.}
  \label{tab:prlm-results-v1}
\end{table}

\begin{figure}[h!]
  \centering
  \includegraphics[width=0.75\textwidth]
                  {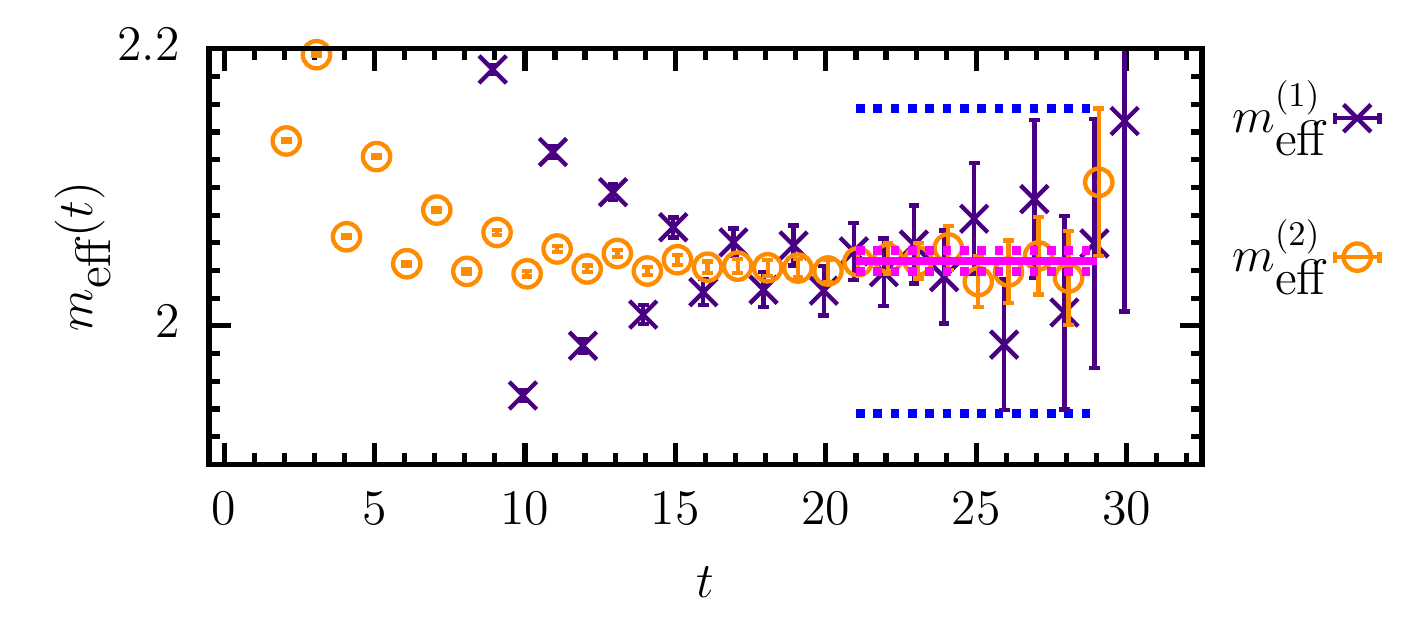}
  \caption{ Effective masses for the 1+0 fit.}
  \label{fig:meff:1+0}
\end{figure}

As explained in Subsection \ref{sec:deter-A0-E0}, we determine the
initial guess for the 1+0 fit using the Newton method.
The fitting results for the 1+0 fit are summarized in the first column
of Table \ref{tab:prlm-results-v1}.
In Fig.~\ref{fig:meff:1+0}, we present results for the effective
masses $m_\text{eff}^{(1)}$ and $m_\text{eff}^{(2)}$, where
\begin{align}
  m_{\text{eff}}^{(j)}(t) &= \frac{1}{j} \ln \left(
  \frac{C(t)}{C(t+j)} \right) \,.
\end{align}
We set the fit range for the 1+0 fit to the region where the effective
mass signal does not oscillate with respect to time.
It corresponds to the magenta color in Fig.~\ref{fig:meff:1+0}.
We set up the Bayesian prior information (info) for 1+1 as follows.
\begin{align}
  A^\text{p}_0 &= A^{[1+0]}_0 \pm [0.8 \times A^{[1+0]}_0]
  \\
  E^\text{p}_0 &= E^{[1+0]}_0 \pm [14.53 \times \sigma^{[1+0]}_{E_0}]
\end{align}
where the superscript ${}^\text{p}$ represent the prior info.
Here, we take the maximum fluctuation of the effective masses within
the 1+0 fit range as the prior width for $E_0$, which corresponds to
the blue dashed line in Fig.~\ref{fig:meff:1+0}.
%

We set the fit range for the 1+1 fit to $13 \le t \le 29$ so that it
minimize the $\chi^2$/d.o.f with a given prior info.
We also apply the same principle of the minimum $\chi^2$/d.o.f to find
the optimal fit ranges for the 2+1 and 2+2 fits.
In the 2+1 fit, we perform the stability test on $A_0$ and $E_0$ to
find the optimal prior widths such that we find the minimum value which
does not change the fit results for $A_0$ and $E_0$.
In the 2+2 fit, we perform the same stability test on $R_1$ and
$\Delta E_1$ to find the optimal prior widths.
In the first 2+2 fit, the fit results for $R_2$ and $\Delta E_2$
shift from the prior info by about 1$\sigma$.
Hence, we update the prior info for the second 2+2 fit to reflect on
this shift.

At present we are working on the 3+2 and 2+3 fits to consume the entire
time slices for the fit range $1 \le t \le 29$. 
%
%

%
\section{Conclusion}
\label{sec:conc-to-do}
%
We have multiple options to choose time slices when we apply the
Newton method to obtain the initial guess.
This provides a natural test to check whether the $\chi^2$ minimizer
finds a local minimum or the global minimum.
In addition, the Newton method reduces the number of iterations
for the $\chi^2$ minimizer dramatically.
At present, the results which we present here are preliminary, but
good enough to insure that the Newton method is highly promising.
Our final results will be available soon.
Please stay tuned for our future report. 

%
%

\acknowledgments

The research of W.~Lee is supported by the Mid-Career Research Program
Grant [No.~NRF-2019R1A2C2085685] of the NRF grant funded by the Korean
government (MOE).
This work was supported by Seoul National University Research Grant
[No.~0409-20190221].
W.~Lee would like to acknowledge the support from the KISTI
supercomputing center through the strategic support program for the
supercomputing application research [KSC-2017-G2-0009,
  KSC-2017-G2-0014, KSC-2018-G2-0004, KSC-2018-CHA-0010,
  KSC-2018-CHA-0043, KSC-2020-CHA-0001].
Computations were carried out in part on the DAVID supercomputer at
Seoul National University.

\bibliography{ref}

\providecommand{\href}[2]{#2}\begingroup\raggedright\begin{thebibliography}{10}

\bibitem{ElKhadra:1996mp}
A.~X. El-Khadra, A.~S. Kronfeld, and P.~B. Mackenzie {\em Phys. Rev.} {\bf D55}
  (1997) 3933--3957, [\href{http://xxx.lanl.gov/abs/hep-lat/9604004}{{\tt
  hep-lat/9604004}}].

\bibitem{Oktay:2008ex}
M.~B. Oktay and A.~S. Kronfeld {\em Phys. Rev.} {\bf D78} (2008) 014504,
  [\href{http://xxx.lanl.gov/abs/0803.0523}{{\tt 0803.0523}}].

\bibitem{Bailey:2020uon}
{\bf LANL-SWME} Collaboration, J.~A. Bailey, Y.-C. Jang, S.~Lee, W.~Lee, and
  J.~Leem {\em Phys. Rev. D} {\bf 105} (2022), no.~3 034509,
  [\href{http://xxx.lanl.gov/abs/2001.05590}{{\tt 2001.05590}}].

\bibitem{Bailey:2017xjk}
J.~A. Bailey, T.~Bhattacharya, R.~Gupta, Y.-C. Jang, W.~Lee, J.~Leem, S.~Park,
  and B.~Yoon {\em EPJ Web Conf.} {\bf 175} (2018) 13012,
  [\href{http://xxx.lanl.gov/abs/1711.01786}{{\tt 1711.01786}}].

\bibitem{Follana:2006rc}
E.~Follana, Q.~Mason, C.~Davies, K.~Hornbostel, G.~P. Lepage, J.~Shigemitsu,
  H.~Trottier, and K.~Wong {\em Phys. Rev.} {\bf D75} (2007) 054502,
  [\href{http://xxx.lanl.gov/abs/hep-lat/0610092}{{\tt hep-lat/0610092}}].

\bibitem{Yoon:2016dij}
B.~Yoon {\em et~al.} {\em Phys. Rev.} {\bf D93} (2016), no.~11 114506,
  [\href{http://xxx.lanl.gov/abs/1602.07737}{{\tt 1602.07737}}].

\bibitem{Bazavov:2012xda}
A.~Bazavov {\em et~al.} {\em Phys. Rev.} {\bf D87} (2013), no.~5 054505,
  [\href{http://xxx.lanl.gov/abs/1212.4768}{{\tt 1212.4768}}].

\bibitem{Bazavov:2011aa}
A.~Bazavov {\em et~al.} {\em Phys. Rev.} {\bf D85} (2012) 114506,
  [\href{http://xxx.lanl.gov/abs/1112.3051}{{\tt 1112.3051}}].

\bibitem{Broyden:1970bro}
C.~G. Broyden {\em {IMA} Journal of Applied Mathematics} {\bf 6} (1970), no.~1
  76--90.

\bibitem{Fletcher:1970fle}
R.~Fletcher {\em The Computer Journal} {\bf 13} (1970), no.~3 317--322.

\bibitem{Goldfarb:1970gol}
D.~Goldfarb {\em Mathematics of Computation} {\bf 24} (1970), no.~109 23--26.

\bibitem{Shanno:1970sha}
D.~F. Shanno {\em Mathematics of Computation} {\bf 24} (1970), no.~111
  647--656.

\bibitem{Press:2007nr}
W.~H. Press, S.~A. Teukolsky, W.~T. Vetterling, and B.~P. Flannery, {\em
  Numerical Recipes}.
\newblock Cambridge University Press, 3~ed., 2007.
\newblock pages 477--483.

\bibitem{Broyden:1965br}
C.~G. Broyden {\em Mathematics of Computation} {\bf 19} (1965), no.~92
  577--593.

\end{thebibliography}\endgroup

\end{document}